\documentclass[journal]{vgtc}                
\ifpdf
  \pdfoutput=1\relax                   
  \usepackage{graphicx}                
  \DeclareGraphicsExtensions{.pdf,.png,.jpg,.jpeg} 
\else
  \usepackage{graphicx}                
  \DeclareGraphicsExtensions{.eps}     
\fi%

\graphicspath{{figures/}{pictures/}{images/}{./}} 

\usepackage{microtype}                 
\PassOptionsToPackage{warn}{textcomp}  
\usepackage{textcomp}                  
\usepackage{mathptmx}                  
\usepackage{times}                     
\usepackage{cite}                      
\usepackage{tabu}                      
\usepackage{booktabs}                  

\usepackage{amsmath} 
\usepackage{amssymb} 
\usepackage{placeins} 

\usepackage[all]{nowidow}
\usepackage[]{siunitx}

\onlineid{0}

\vgtccategory{Research}
\vgtcpapertype{please specify}

\title{AccuStripes: Adaptive Binning for the Visual Comparison of Univariate Data Distributions}

\author{Anja Heim, Eduard Gröller, and Christoph Heinzl}
\authorfooter{
\item
 Anja Heim is with University of Applied Sciences Upper Austria and TU Wien. E-mail: anja.heim@fh-wels.at.
\item
 Eduard Gröller is with TU Wien and VRVis. E-mail: groeller@cg.tuwien.ac.at.
\item
 Christoph Heinzl is with University of Passau. E-mail: christoph.heinzl@uni-passau.de.
}

\shortauthortitle{Heim \MakeLowercase{\textit{et al.}}: AccuStripes}

\abstract{ Understanding and comparing distributions of data (e.g., regarding their modes, shapes, or outliers) is a common challenge in many scientific disciplines. Typically, this challenge is addressed using side-by-side comparisons of histograms or density plots. However, comparing multiple density plots is mentally demanding. Uniform histograms often represent distributions imprecisely since missing values, outliers, or modes are hidden by a grouping of equal size. In this paper, a novel type of overview visualization for the comparison of univariate data distributions is presented: AccuStripes (i.e., “accumulated stripes”) is a new visual metaphor encoding accumulations of data distributions according to adaptive binning using color coded stripes of irregular width. We provide detailed insights about challenges of binning. Specifically, we explore different adaptive binning concepts such as Bayesian Blocks binning and Jenks’ Natural Breaks binning for the computation of binning boundaries, in terms of their capabilities to represent the datasets as accurately as possible. In addition, we discuss issues arising with the representation of designs for the comparative visualization of distributions: To allow for a comparison of many distributions, their accumulated representations are plotted below each other in a stacked mode. Based on our findings, we propose three different layouts for comparative visualization of multiple distributions. The usefulness of AccuStripes is investigated using a statistical evaluation of the binning methods. Using a similarity metric from cluster analysis, it is shown, which binning method statistically yields the best grouping results. Through a user study we evaluate, which binning strategy visually represents the distribution in the most intuitive form and investigate, which layout allows the user the comparison of many distributions in the most effortless way.
\\%
} 

\keywords{Adaptive binning, Visual comparison, Univariate distributions.}

\CCScatlist{ 
 \CCScat{K.6.1}{Management of Computing and Information Systems}%
{Project and People Management}{Life Cycle};
 \CCScat{K.7.m}{The Computing Profession}{Miscellaneous}{Ethics}
}

\teaser{
  \centering
  \includegraphics[width=\linewidth]{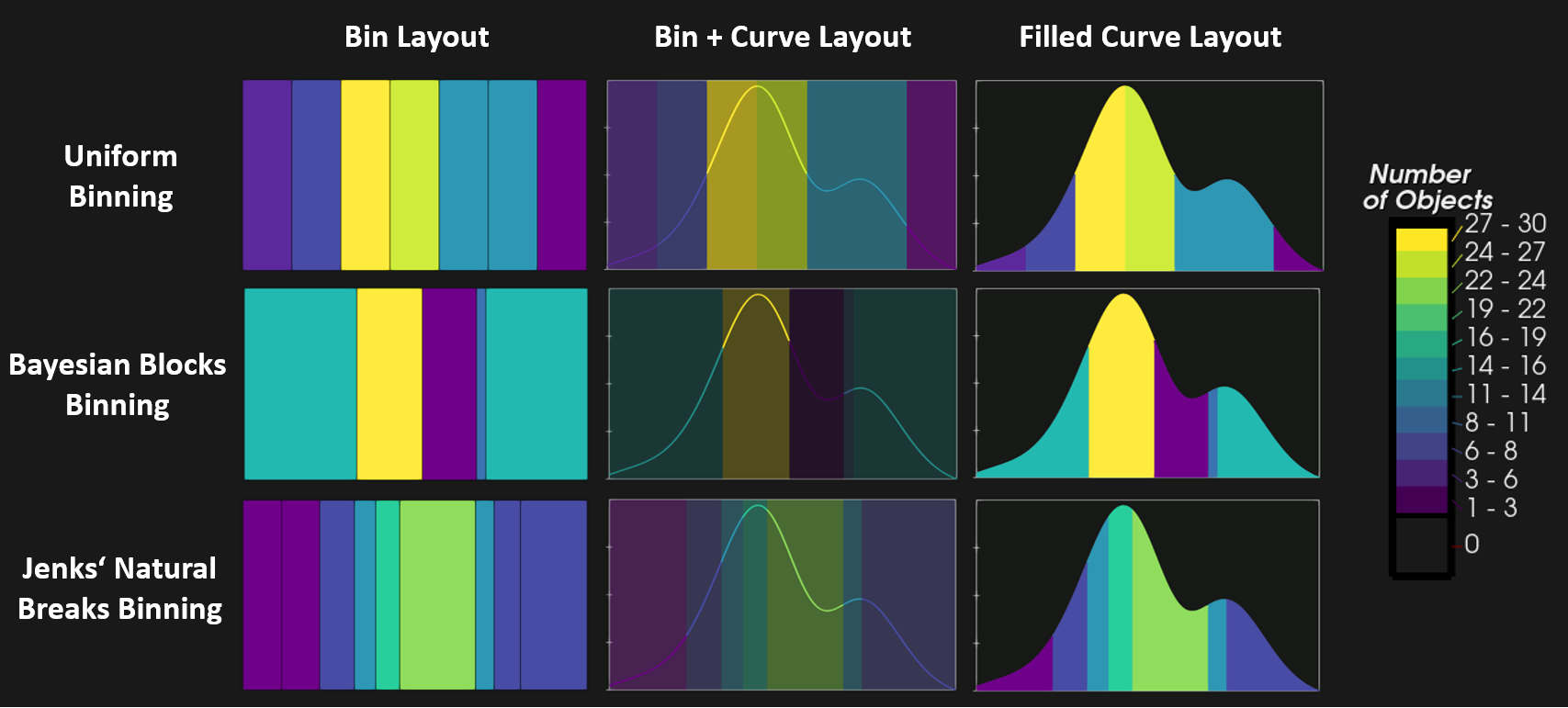}
  \caption{AccuStripes: A new visual methaphor to represent univariate distributions to ease structural estimation and comparison tasks. AccuStripes may be presented in nine different variants, resulting from the composition of three binning methods (uniform binning, Bayesian Blocks, Jenks' Natural Breaks) and three different layout strategies (Bin Layout, Bin+Curve Layout, Filled Curve Layout).}
  \label{fig:teaser}
}

\vgtcinsertpkg

\begin{document}


\firstsection{Introduction}

\maketitle
The analysis of \textbf{univariate, continuous distributions} is essential to gain new insights in many scientific and technical domains, such as economics, engineering or materials science. For instance, material experts need to identify and compare distributions of attributes, such as the characteristics of fibers (e.g., length, orientation, position, etc.) in fiber reinforced composites, to determine the properties of material samples, and ultimately, at which load a fracture occurs \cite{MAURER2019}. To get an understanding of the shape of the distributions and their relation to each other, analysis tasks as \textbf{structural estimation} and \textbf{comparative investigations} are essential \cite{Blumenschein2020a}. Structural estimation tasks include examining the shape and type of a distribution and determining flaws, such as gaps, outliers, or spikes. Comparative tasks focus on matching the similarity of distributions through comparing their shape by identifying the location of modes, spikes or outliers \cite{Correll2019}.

A plethora of visualization methods exists for analyzing distributions \cite{Blumenschein2020a, Thrun2020}. Shape-based charts show the entire distribution, but are difficult to compare side by side if there exist a multitude of plots. Histogram-based charts generally build upon a uniform and easy-to-compare summarization of the data, but are highly susceptible to visualization mirages \cite{Correll2019, McNutt2020}. Uniform binning techniques take little to no account of the underlying data when forming bins, thus imprudent settings of design parameters, such as bin width, lead to crude representations of the distributions, e.g., hiding missing data.

To support the visual analysis of data distributions, we developed a new visual metaphor called "AccuStripes" (see \autoref{fig:teaser}). We designed AccuStripes with two intentions in mind: (1) presenting the shape of distributions with higher accuracy to increase the chance of detecting structures as gaps, outliers or spikes, and (2) allowing users to perform tasks of structural estimation and comparative analysis of distributions with equal efficiency. AccuStripes are a visual representation of an accumulation of a single distribution resulting from the application of an adaptive binning technique. The individual bins are visualized through a color coded stripe of variable width. The juxtapostioning of all bins results in the AccuStripes representation. To compare several distributions, each distribution is represented as AccuStripes and arranged one below the other. Our prototypical implementation of AccuStripes is based on the overview visualization of CoSi~\cite{Heim2021}.
The main contributions of this paper are found in the following aspects:

\begin{enumerate}
\item Based on a detailed investigation regarding the challenges of binning techniques, we examine \textbf{two different adaptive binning techniques: Bayesian Blocks and Jenks' Natural Breaks binning}. These methods perform an accumulation of a distribution by considering the location of the underlying data points. Therefore, a more precise aggregation of the data can be achieved.

\item We explore three different layout designs to ease the visual comparison of multiple univariate distributions: The \textbf{Bin layout} represents each distribution merely by rectangular bins. The \textbf{Bin + Curve layout} displays distributions using bins as well as their curved form. Finally, the \textbf{Filled Curve layout} shows the data merely by their curved form, but the area under the curve is colored by the selected binning technique.

\item Through combining a \textbf{statistical evaluation} of the binning methods and a \textbf{user study}, we determine if adaptive binning techniques accumulate the shape of distributions more intuitively for observers than uniform binning. We also examine which layout design is more suited to compare many distributions.
\end{enumerate}

The remainder of this paper is organized as follows. In \autoref{sec:Background} we give an overview of the mathematical background for the analysis of distributions and discuss related visualisation techniques in \autoref{sec:RelatedWork}. A detailed description of the reasoning behind AccuStripes and their design choices are given in \autoref{sec:Method}. In \autoref{sec:Evaluation} we provide information about our statistical evaluation and user study. Finally, we discuss how AccuStripes deal with flaws in the data in \autoref{sec:Discussion} and conclude in \autoref{sec:Conclusion}.

\section{Background} \label{sec:Background}
The analysis of univariate, continuous datasets is an essential challenge in many disciplines. In this section we give an overview of analysis tasks performed on univariate data distributions and why we focus specifically on identifying and comparing distributions. Furthermore, we give an overview of methods for the aggregation and density estimation of univariate distributions and explain why a purely analytical analysis alone is often not sufficient for the comparison of distributions.

\subsection{Analysis Tasks for Distributions}
Szafir et al. \cite{Szafir2016a} and Blumenschein et al. \cite{Blumenschein2020a}, both give a detailed explanation of analysis tasks for univariate data distributions. The goal of our work is to accurately represent the fine shape of distributions, while at the same time abstracting them to the point where effortless comparison is achievable. The structures of the distributions we are interested in are their shapes, i.e., their uniformity, bi- or multimodality, whether the tail of the distribution is truncated or not, and whether the distribution has a strong skewness or is normally distributed \cite{Thrun2020}. The analysis tasks we strive to support are structural estimation, as defined by Szafir et al.\cite{Szafir2016a}, or global tasks, as determined by Blumenschein et al. \cite{Blumenschein2020a}: Both, structural estimation as well as global tasks, specify tasks of observers as recognizing patterns in distributions, that cannot be intuitively captured by individual statistics, e.g., describing the shape, type, or skewness. The comparison tasks belong to these categories as well.
We also aim to detect flaws in the data, such as gaps, i.e., small areas of missing data, outliers, and peaks, i.e., strong and small accumulations of data points \cite{Correll2019}. In our work, we do not include summarization tasks, since aggregated statistical measures usually require basic knowledge of statistical methods and their interpretation. Moreover, they can be misleading, as can be seen in the Anscombe's quartet \cite{Anscombe1973}, in which different distributions are described by identical aggregated characteristics. Although we target to enable the identification of outliers, spikes and gaps, our goal is not to guarantee exact identification tasks with precise value specifications, such as minimum or maximum. Our priority is to allow observers to determine the presence and an approximate location of flaws and to match these observations with other distributions.

\subsection{Density Estimation for Univariate Distributions}
Density estimation is the construction of an estimate of the density from the available data points \cite{Silverman1986}. The histogram and the kernel density estimator are both methods that compute through different approaches the density of continuous data distributions.

The histogram provides a subdivision of data points into non-overlapping groups, so-called bins, based on predefined origins, numbers of bins or interval sizes. There are many different binning strategies, most of which are used in image processing, computer vision, and machine learning. In general, there are two different strategies for grouping data: uniform and adaptive binning. In uniform binning all bins have the same width. The width or number of bins can be estimated by rules of thumb, such as Sturge's rule or Scott's normal reference rule, but there exists no method that can determine the best number of bins for every type of data \cite{Sahann2021}. 

Adaptive binning in contrast produces variable-sized bins that are calculated based on the underlying data. The simplest adaptive binning method is equal-frequency binning \cite{Brewer2002}, where each bin contains the same number of data points. For this reason the position of the bin boundaries is arbitrary. A more sophisticated adaptive binning technique is Bayesian Blocks binning (BB) \cite{Pollack2017}, where the bin widths are variable and bin boundaries are determined by the data distribution. The concept of BB is to approximate the distribution as partially constant and to find the optimal number of constant ranges (i.e., blocks or bins) and their positioning. Each block or bin is separated by alteration points at the boundaries. The bins start with the first data point and end with the last data point without gaps, although the bins can be empty. For the computation, each bin is described through a fitness function that depends on the location and number of alteration points. It is further optimised to find the correct number and positioning of points. The fitness function can be maximized by varying the number of data points in each bin and the width of each bin. The overall fitness function is the sum of the fitness functions of the individual bins. The overall fitness function is thus computed $n-1$ times for each of $n$ bins, resulting in a computation time of $O(N^{2})$ \cite{Pollack2017}.

Clustering can also be seen as an adaptive binning technique, as this technique also divides data into groups of different sizes. For one-dimensional distributions, the Jenks' Natural Breaks Classification method (NB), also referred as Fisher-Jenks algorithm, is particularly suitable, since it divides the data points into groups based on their arrangement. NB seeks to optimally classify $n$ data points into $k$ classes, or bins, such that inter-class variance is maximized and intra-class variance is minimized, which has time complexity of $O(k\cdot n^{2})$ \cite{Fisher1958, Jenks1977}. We rely on an improved version of the original implementation as presented by Hilferink \cite{Hilferink_2021}, which features a time complexity of $O(k\cdot n \cdot log(n))$ through minimizing the sum of the squared deviations from the class means. NB requires the user to provide a number of $k$ classes prior to executing the algorithm. To obtain a suitable number of $k$ classes, the NB algorithm is started with a small $k$ and its resulting binning can be quantified through the goodness of variance fit (GVF). The GVF computes the sum of deviations within the classes, resulting in a value between (0) (worst fit) to 1 (perfect fit) \cite{Fariza2016}. In each iteration the number of classes is increased until a minimal GVF is reached.

Regardless of the method used to determine the bins, the histogram suffers from its main purpose of binning the data. The data points are deprived of their individual positions by replacing them with an aggregated bin position, making the shape of the histogram discontinuous and flat in each bin. Kernel Density Estimation addresses this issue by calculating a smooth density estimate \cite{Weglarczyk2018}. The kernel estimator of the kernel density estimation calculates for each individual data point an interval of many data values, which are distributed around the respective data point. This computation results in an estimate of the density from the available data. There are various density estimators existing in literature, however, the choice of kernel function has less impact on the shape of the curve compared to the impact of the bandwidth. A commonly used method for the computation of the bandwidth is Silverman's rule of thumb \cite{Silverman1986}, as it is easy, fast, and robust to compute.
        
\subsection{Comparing Distributions}
Distributions can be compared using two different methods: vectorial or probabilistic. If the distribution is interpreted as a vector, geometric distances (e.g., Minkowski distances) can be applied. In probabilistic methods, the overlap of the two distributions to be compared is measured. Cha \cite{Cha2007} lists a variety of different measures. Making a suitable choice from the huge variety is challenging for non-statisticians. To compare histograms, specialised distance measures exist, which can be classified as bin-to-bin or cross-bin distances \cite{Ma2010,Bazan2019}. Since bin-to-bin methods only measure the difference between the bins of the same interval, slight shifts in the histograms can lead to different results. In addition, bin-to-bin methods are sensitive to changes in the bin width. Comparing the same histograms in different binning modes can lead to differing results. The current problem with cross-bin measures is that they often lead to worse results than bin-to-bin methods, as they overestimate the correlations between the bins \cite{Bazan2019}. Picking an appropriate method can be complex and even experts in the field of statistics admit that certain data characteristics are easier to detect and analyse through visual representations \cite{Laeuter1988,Rodrigues2019}. To fulfil the task of comparing univariate distributions, we consider visual inspection to be as important as analytical examination.

\section{Related Work} \label{sec:RelatedWork}

\subsection{Visualization Techniques for Distributions} 
The variety of visualization techniques for distributions is manifold and can be found in the visualization of statistical data and in time series visualization. Blumenschein et al.~\cite{Blumenschein2020a} give a detailed overview of the most common forms of chart designs. Representations depicting the shape of distributions are characterised by histogram- and shape-based representations. Hybrid forms combining these two types are also used. In histogram-based charts, the distributions are grouped and plotted by bars or lines, where the height represents the size of the groups. The bars can be arranged in a large variety of ways. They can be plotted horizontally or vertically. Bars belonging to different distributions can be plotted next to each other (mirrored bar chart), on top of each other (stacked bar chart), or alternately (grouped bar charts). All these variants have in common that the observer has to compare the height of the individual bars, i.e., the position and length of the boxes, in order to be able to evaluate the distribution or its comparison. In shape-based charts, the distribution is plotted as a density using a continuous line (line chart), whose area under the curve can be filled (area chart). The height of the line indicates the frequency of the data points at a specific location. Respective representations also occur in many different arrangements and variants. The densities can be plotted one below the other, superimposed, or mirrored (violin plot, bean plot). These specific variants are particularly useful for the direct comparison of two distributions, as observers can accurately compare the location and height of the curves. The depiction of multiple distributions in a matrix had also been experimented with, yet turned out to be mentally challenging for the viewers~\cite{Blumenschein2020a}. Another technique for visualising distributions are dot plots~\cite{Correll2019}, where each data point is displayed as a point with a certain transparency.

Recently, studies have been conducted that address the issue of determining which of these designs is better suited for which analytical task. Rodrigues et al.~\cite{Rodrigues2019} evaluated the effectiveness of different diagrams for characterising distributions and for comparing them. Histograms performed best for representing distributions, while bar and line charts were most effective for comparative tasks. In the evaluation of Blumenschein et al. \cite{Blumenschein2020a}, bar charts and density charts placed one below the other performed best in identifying and also comparing the shape of distributions. Heatmaps were not considered in these studies. Hybrid charts, as the Summary Plot by Potter et al.~\cite{Potter2010} or the V-plot by Blumenschein et al.\cite{Blumenschein2020a}, combine several designs to support different analysis tasks at the same time, resulting in a sophisticated representation.

\subsection{General Aspects of Aggregation Techniques in Visualization}
The visualization of large quantities of data is closely related to scalability challenges. Since the human perceptual system is a limiting factor for processing information, aggregation is essential to present data in an accessible way. Data should be shown in an abstracted form so that the available screen real estate is optimally exploited and the entire data can be grasped at a glance~\cite{Albers2011}. Respective techniques in visualization can be found in many variants, e.g. point, line, and area representations are common. Kehrer et al.~\cite{Kehrer2013} visualize categorical time series data by aggregating temporal changes with tiny bar graph glyphs. Their color coding and positioning in a side-by-side style allow for a comparison of the entire data. Through an overview visualization technique called Colored Mosaic Matrix, the authors Kobayashi et al.~\cite{Kobayashi2013} visualize high-dimensional categorical meteorological data through colored rectangles arranged in a tabular form. Especially in small drawing areas, this technique achieved a high degree of legibility among observers. Sequence Surveyor~\cite{Albers2011} is a table-based visual representation where patterns in genome data can be effectively detected using color. These works show that aggregations allow observers to perceive complex data in a comprehensible form easing the task of comparison. Since each of these techniques was developed for its respective type of data, a representation of univariate distributions is not feasible.

Saito et al.~\cite{Saito2005} accumulated the shape of univariate distributions through a two-tone coloring and displayed the data as density plots underneath each other. Although the design is highly space-saving, datasets with severe alternations are difficult to compare. CoSi~\cite{Heim2021} is a histogram-based heatmap representation for the comparison of multiple distributions. The binning of distributions was performed uniformly. Uniform binning suffers from the drawback that the aggregation of data is performed in a predetermined manner without considering the underlying data distribution, thus potentially leading to visualization mirages. As defined by McNutt et al.~\cite{McNutt2020}, visualization mirages are defined as visualizations that convey assumptions to the viewer that, upon closer inspection, lack information because it is not obviously recognizable. 

\section{AccuStripes} \label{sec:Method}
AccuStripes are a visual metaphor to represent a distribution in an accumulated style. In our work we define a distribution to be an univariate, continuous vector $ \vec{x}= \begin{bmatrix} a_{1}, & a_{2}, & \cdots & , a_{n} \end{bmatrix} $, where $a_{i}$ are the individual data points and $ a_{i} \in \mathbb{R} $. The number of data points $n$ to be visualized by AccuStripes is only limited by the runtime of the BB binning. We tested distributions with a maximum of $100,000$ points. AccuStripes is the advancement of the visualisation framework CoSi \cite{Heim2021}, which was designed for the comparative display of datasets from in-situ tests of materials. These tests usually produce five to ten datasets to be compared. For this reason, AccuStripes was developed to focus on the comparison of a number of datasets in the range of five to ten. Our framework implementing AccuStripes is available on github as part of the application framework open\_iA \cite{Froehler2019}.

\begin{figure*}[tbp]
  \centering
  \mbox{} \hfill
  \includegraphics[width=\linewidth]{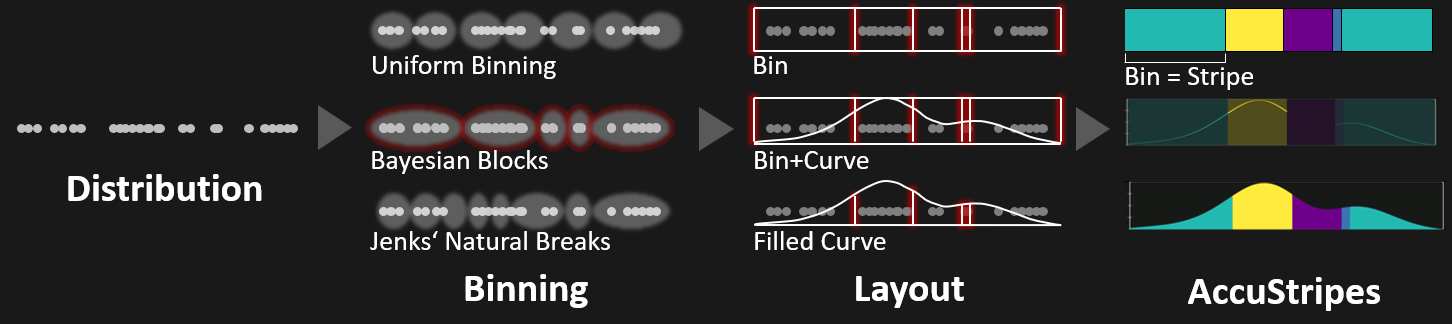}
  \hfill \mbox{}
  \caption{\label{fig:infographic}%
    Generation of AccuStripes: First the distribution is binned with a method selected by the user. Here, BB binning was chosen, indicated in red. Then three different layouts are prepared for display. The user can switch between the different layouts.}
\end{figure*}

\autoref{fig:infographic} summarizes the process of creating AccuStripes. AccuStripes follow the idea that the representation of distributions should match their shape closely, but still allow for easy comparison. Visual matching by observers ought to be feasible with little mental effort, thus the distributions should be accumulated. Adaptive binning methods met both our requirements. Grounded by an intensive literature research, and elaborated in more detail in the following \autoref{subsec:binning}, BB and NB binning were chosen to aggregate the data. The user determines which of these two methods will be applied to the distribution. The selected binning method produces a specific number of bins with variable width and all containing a varying number of data points. For the mapping of the numerical values of the data points into a visual encoding, we transform the values of the distribution and the bin boundaries in our visual workspace into the interval $[0,1]$. For each bin a color coded stripe is drawn from the bin boundary of the previous bin to its own boundary. The color of the stripe encodes the number of data points contained in the bin. For coloring the color scheme \textit{Viridis} is applied \cite{Liu2018}. The horizontal juxtaposition of all stripes according to their bin boundaries creates our visual metaphor, the AccuStripes. Thus, a single distribution is mapped to AccuStripes. For the comparison of several distributions, all distributions are first brought to a common range. Then the adaptive binning method is applied on each distribution individually. Next, the distributions are transformed into our visual workspace, where for each of them their own AccuStripes representation is created. In a last step, the individual AccuStripes are positioned one below the other in a stacked mode. 

The design we have just described is called \textit{Bin layout}. This layout displays the bins based only on stripes, or rectangles, of the same height but varying widths. As discussed in \autoref{subsec:layout}, there is no clear evidence whether visual encodings of bars or lines are more effective in visual comparison. For this reason, we offer observers two additional layouts, beside the Bin layout, between which they can switch. The second layout that will be displayed to the users after the Bin layout, is the \textit{Bin + Curve layout}. As the name suggests, this layout displays both the rectangular stripes and a line representation of the distribution simultaneously. The third choice called \textit{Filled Curve layout} concentrates on the line representation of the distribution. Here, the attention of the user is focused on the curve shape of the distribution and the layout displays the binning result solely in the area under the curve.

To evaluate the usability of the adaptive binning methods, they are compared to uniform binning. The number of bins is determined by the Sturge’s rule, since it was the only method that determined a small number of bins ($\sim20$) for datasets consisting of $\num{1000000}$ data points in the evaluation of Sahann et al.~\cite{Sahann2021}. In the demonstrative implementation of our new method, the uniform binning can also be shown in all three layouts.

\begin{figure*}[h!tb]
  \centering
  \mbox{} \hfill
  \includegraphics[width=\linewidth]{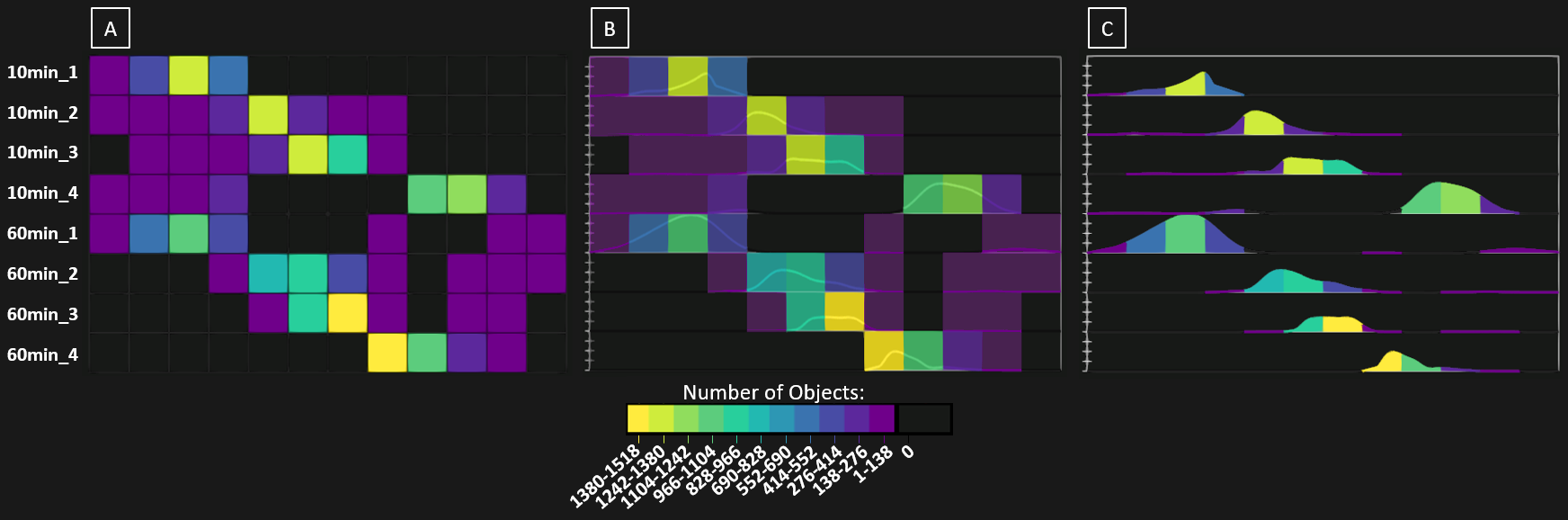}
  \hfill \mbox{}
  \caption{\label{fig:UseCase_UB}%
    The AccuStripes were generated using \textbf{uniform binning}: A) \textit{Bin layout}, B) \textit{Bin + Curve layout}, and C) \textit{Filled Curve layout}. Similar color patterns indicate similar distributions, but this binning does not result in a clearly apparent pattern. }
\end{figure*}

\begin{figure*}[h!tb]\textbf{}
  \centering
  \mbox{} \hfill
   \includegraphics[width=\linewidth]{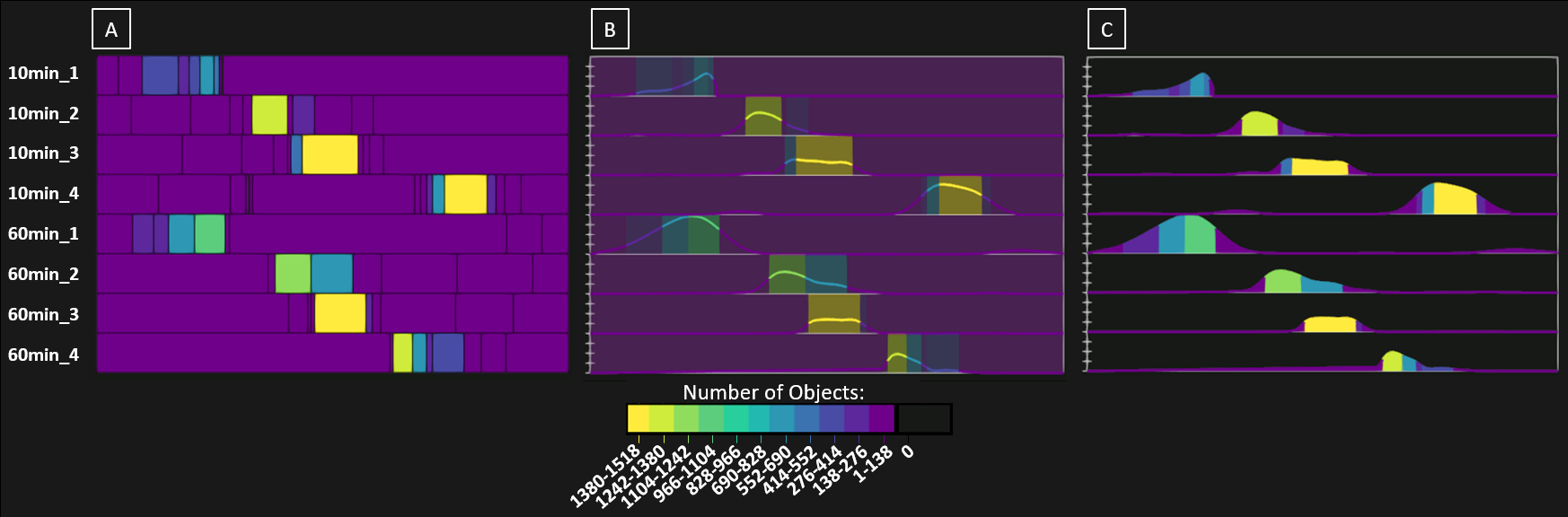}
  \hfill \mbox{}
  \caption{\label{fig:UseCase_BB}%
    The AccuStripes were generated using \textbf{BB binning}: A) \textit{Bin layout}, B) \textit{Bin + Curve layout}, and C) \textit{Filled Curve layout}. The most similar distributions, i.e. those with the same indices as $10min\_3$ and $60min\_3$, exhibit the most similar color patterns.}
\end{figure*}

\begin{figure*}[h!tb]
  \centering
  \mbox{} \hfill
   \includegraphics[width=\linewidth]{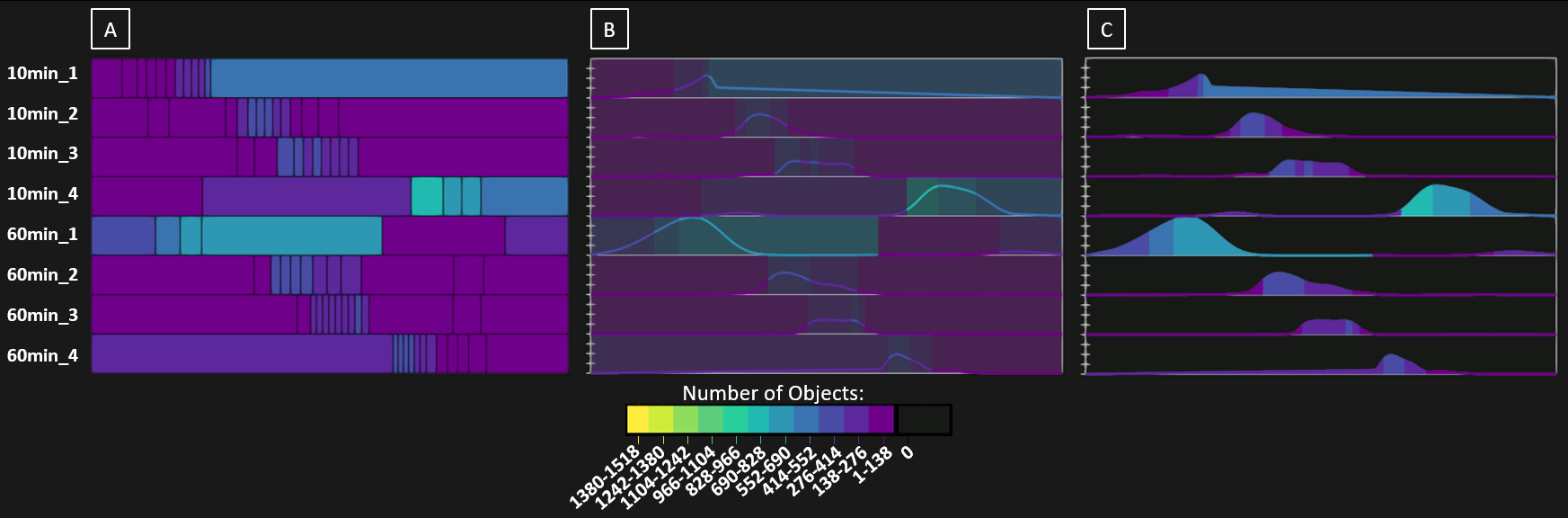}
  \hfill \mbox{}
  \caption{\label{fig:UseCase_NB}%
     The AccuStripes were generated using \textbf{NB binning}: A) \textit{Bin layout}, B) \textit{Bin + Curve layout}, and C) \textit{Filled Curve layout}. Compared to the uniform binning of \autoref{fig:UseCase_UB}, this adaptive binning also produces better perceptible patterns. This binning leads to a finer grouping of the data points, thus a smaller number of points is contained in all bins. As a result, the color variation is significantly lower compared to BB and uniform binning.}
\end{figure*}

\subsection{Challenges of Binning Techniques} \label{subsec:binning}
Uniform binning is a standard method for aggregating distributions. It is due to the fact that binning is a very simple, efficient and fast to compute method that only requires counting how many elements fall into a predefined interval, i.e., a bin. In general, the details are recorded in a histogram, even if they are not necessarily visualized. Moreover, if elements constitute only a small subset of a dataset, this aspect is captured in the histogram and is not suppressed by the more dominant groups of elements. Another benefit of uniform binning is that dominant groups of elements cannot be overrepresented. If the largest portion of a distribution consists of a single group of elements, all elements will still be grouped into a single bin. Furthermore, the comparison of uniformly binned representations is consistent. Since the number of bins in comparative analyses is typically identical and the datasets are structured in common intervals, the bins in the different datasets have the same bin boundaries. A serious drawback of uniform partitioning of distributions are the artificial boundaries that are introduced: To capture the precise shape of the data distribution, the number of bins must be chosen conscientiously. Sahann et al.~\cite{Sahann2021} evaluated which rules of thumb calculate an acceptable number of bins based on human perception. Their results showed that above a set of $\num{1000}$ data points, Sturge's rule was the only one that did not overestimate the necessary number of bins. The boundaries are defined without considering the underlying spread of the data, which can cause elements actually belonging together to be separated into different bins \cite{Leow2004}. Thus, uniform binning imposes an artificial constraint on the data.

Adaptive binning methods in contrast are able to calculate the bins based on the underlying positioning of the data points. According to the selected method, the number of bins is determined empirically or by rules of thumb, e.g. by k-means clustering or NB binning. However, there are also methods that are capable of calculating the number of bins directly based on the data, such as BB binning. Using these methods, it is possible to cluster irregular variations more accurately than when using uniform group sizes. Filtering out dominant groups in the data is also easier, as noise is more likely to be suppressed. In the case of different datasets consisting of many highly similar data points, adaptive binning performs better at highlighting subtle differences in the datasets, while uniform binning would produce the same results. In general, adaptive methods provide good results in terms of accuracy, low number of bins, and efficient computation time.

We applied the following criteria to determine suitable adaptive binning algorithms for our work: The most important requirement was that the adaptive methods should require no or very few input parameters. Furthermore, the binning results should not be overly sensitive to the choice of these parameters. In addition, we focused on adaptive binning techniques that are applicable and suitable for datasets of any shape. Therefore, clustering algorithms which are merely suitable for convex datasets, such as k-means, were discarded \cite{Xu2015} in our considerations. An adaptive binning method meeting our requirements is BB binning~\cite{Scargle2013}. It is free of input parameters and provides a quantitatively optimal number of bins \cite{Pollack2017}. Based on the computation model of BB binning, the bins have the following properties: Segments of distributions in which the data points are sparse result in larger bins, while segments with a high number of data points are grouped in smaller bins. Furthermore, if the distribution alternates slowly, the bins are wide, and if the values change rapidly, the bins are narrower. In general, each bin contains a "flat" segment of the distribution, i.e., a segment in which no significant alterations of the data values occur. BB binning therefore suppresses effects of minimal fluctuations in the distribution and sets the bin boundaries at positions of significant alteration in the data. BB gives more detailed results when the viewer's interest is higher for peaks than tails. If tails have little, near constant values, BB binning will compute much broader bins, which has the advantage that the average value of these regions becomes easier perceivable as if visualized with uniform bins. For the same reason, BB is more applicable for larger datasets, since in small datasets the alterations can be too small to compute strong alteration points. A clustering algorithm similar to k-means is NB binning \cite{Jenks1977}. In contrast to k-means, NB binning was developed specifically for one-dimensional data. This technique detects groups after sorting the data values. Bin boundaries are created to group similar data points, i.e., with a small difference in their value, most effectively while maximizing the differences between bins. Bin boundaries are set at positions where relatively big alterations or gaps in the data values occur. NB is known to perform good on multi-modal distributions, but is not recommended for data that has a low variance.

\subsection{Challenges in Visualization Design} \label{subsec:layout}
The layout structure of AccuStripes was developed with the aim of comparing several distributions in the order of five to ten. An evaluation of suitable encodings for performing a visual comparison was carried out by Ondov et al. \cite{Ondov2019} and Rodrigues et al. \cite{Rodrigues2019}. Ondov et al. tested pairwise representations of bar charts, line charts, and donut charts in different arrangements such as mirrored, superimposed, or overlaid. In pairwise comparisons line charts performed better than bar charts. Rodrigues et al. identified bar charts as most effective for visual comparison tasks, followed by line charts. Area charts, i.e., line charts in which the area under the curve is filled, performed very poorly for comparison tasks. Gogolou et al. \cite{Gogolou2019} investigated whether line charts, horizon graphs, or colorfields are better suited for comparing time series data. The study found that in line charts and colorfields, the exact height values of the lines are less important, as people tend to focus on the relative values and overall shape. Similarly as Correll et al.~\cite{Correll2012} and Albers et al.~\cite{Albers2014}, Gogolou et al. confirmed color variations as particularly effective in identifying higher-level patterns, like spikes. In this study the participants initially focused on the color to estimate similarity. Only for a detailed comparison, they considered the exact shape and position in the diagrams. In particular, colorfields assist in the recognition of related regions based on similar colors.

Since there is no clear evidence whether rectangular bars or lines are more appropriate, we employ both types of encodings. We decided to test both encodings against each other, which is why the \textit{Bin layout} emphasizes rectangles and the \textit{Filled Curve layout} focuses on lines. We also wanted to investigate if the two encodings together would provide the best representation, which is why we added the \textit{Bin + Curve layout} to our design. We use the Kernel Density Estimation to determine the line based representation of the distributions. The bandwidth is calculated using the commonly employed Silverman's rule of thumb \cite{Weglarczyk2018}. The comparison of different distributions using AccuStripes is primarily based on the visual coding in colors. For the color scheme, we chose the so-called \textit{Viridis} color theme \cite{Liu2018}. Following this theme in AccuStripes, bins with a small number of data points are blueish. As the number of data points in the bins increases, the color of the bins becomes brighter and more yellowish. \textit{Viridis} was introduced as default colormap for matplotlib \cite{Liu2015} and is a multi-hue colormap with monotonically increasing luminance and a smooth arc through blue, green, and yellow hues. Liu et al. \cite{Liu2018} found that multi-hue colormaps, especially \textit{Viridis}, provide improved discrimination, while preserving perception of order. \textit{Viridis} is characterized by its wide perceptual range. It exploits the available color space as much as possible while maintaining uniformity minimizing contrast effects. We decided to use a discrete form of \textit{Viridis} to ease the comparison task, as varying shades of color given by a continuous color map would increase the complexity. Further studies are required to determine whether the error caused by quantification of the colors is more severe than the observer's inability to distinguish and compare colors \cite{Kelly2017}. We wanted to provide users with detailed insights into the data by using multiple color variations in the color scheme. It has been shown that only about 12 different colors can be effectively distinguished from each other, which is why we opted for this number of colors \cite{Munzner2014}. AccuStripes were initially inspired to represent distributions gained in the field of materials science. For experts in this field it was important to recognize empty bins at a glance. Therefore, empty bins were assigned their own color, the background color, to differentiate them from bins containing values.

\subsection{Use Case - Real-World Material Data Distributions} \label{subsec:useCase1}
We demonstrate AccuStripes on a real-world datasets from the domain of materials science. In our example, material experts have to explore the changes in fiber materials after a stress test. As indicated in the report by Heinzl et al. \cite{Heinzl2017}, visual computing can be of great help for this purpose. Recently, Heim et al. \cite{Heim2021} developed a method for representing the change in fibers using univariate distributions of similarity values. Each fiber is represented by a similarity value, which is in the range of $[-0.5,0.5]$. The datasets presented here describe a material that was subjected to a stress test for ten minutes ($10min\_x$) and sixty minutes ($60min\_x$). To ensure an accurate analysis of the material, it has been subdivided into four regions (indicated with index $\_1$ to $\_4$). In summary, eight distributions are compared, each containing approximately $\num{2500}$ fibers, i.e., data values. 

After loading the datasets, our prototype generates a representation for the user as shown in \autoref{fig:UseCase_UB}A. Each of the eight distributions is represented using an AccuStripes metaphor, which are then stacked. We first applied uniform binning to the data. In \autoref{fig:UseCase_UB}A the user can inspect the data in the \textit{Bin layout}. In this layout, the user perceives segments more dominantly in which no data values exist. To examine the distributions in more detail, the user can switch to the \textit{Bin + Curve layout} shown in \autoref{fig:UseCase_UB}B. This layout design supports the user in exploring data segments, in which only a few data values are located. In contrast to the \textit{Bin layout}, the user is able to see the shape of the distribution and can follow the ascent and descent of the curve. If the curve is low and constant, the colored background supports the user in identifying these segments. Next, the user is able to view the \textit{Filled Curve layout} displayed in \autoref{fig:UseCase_UB}C. This layout facilitates the comparison of the modes in the distributions. Segments with few data points are pushed into the background due to their small size. Then, the user can examine the AccuStripes computed by the BB binning seen in \autoref{fig:UseCase_BB}. Again, \autoref{fig:UseCase_BB}A displays the \textit{Bin layout}, \autoref{fig:UseCase_BB}B shows the \textit{Bin + Curve layout}, and \autoref{fig:UseCase_BB}C demonstrates the \textit{Filled Curve layout}. In contrast to the result of uniform binning, various structures are recognizable through the BB binning. For example, the user can clearly notice that the datasets with the same indices, e.g., $10min\_3$ and $60min\_3$, are most similar to each other based on their color pattern. Lastly, the user can inspect the NB binning, illustrated in \autoref{fig:UseCase_NB}. \autoref{fig:UseCase_NB}A shows the \textit{Bin layout}, \autoref{fig:UseCase_NB}B demonstrates the \textit{Bin + Curve layout}, and \autoref{fig:UseCase_NB}C displays the \textit{Filled Curve layout}. With this adaptive binning method, the user can also detect similar color patterns between distributions with the same indices. Based on this use case, we can see that adaptive binning techniques can be useful to identify similar distributions from a given number of datasets.

\section{Evaluation} \label{sec:Evaluation}
It was important to us to represent the distributions as accurately as possible in an accumulated form, yet, a simple, user-friendly interpretability of the resulting representation has been of utmost priority. For this reason, we focused on evaluating the design readability of AccuStripes based on the applied binning and layout strategy for comparison. Furthermore, we were interested whether the statistically superior binning method also provided the best comprehensibility for observers. Therefore, we first conducted a quantitative evaluation of the binning methods. Our main focus lay on the user study, following the design considerations by Sedlmair et al. \cite{Sedlmair2012}, where we investigated which binning procedure yielded perceptually better results and which layout design supported participants in solving comparison tasks.

The shape of distributions can be evaluated by various structures, such as the number of modes, whether it is skewed or normally distributed, whether it belongs to a certain family of distributions, etc. \cite{Silverman1986}. To ensure a systematic evaluation of AccuStripes, we examine distributions generated by the procedure described by Correll et al. \cite{Correll2019}: All our distributions are generated from Gaussian distributions of three different size categories: using $\num{1000}$, $\num{10000}$, and $\num{100000}$ data points. The distributions were randomly generated in the range of $[0,1]$. We refer to the data points contained in Gaussian distribution as correct. To create distributions of varying shapes, we added so-called erroneous data points to or removed correct ones to the extent of $0\%-25\%$ from the original data points. Beside adding data points to generate noise and a skewness, we focus on three different types of structures: gaps, outliers, and spikes. Gaps are caused by removing points at a certain location in the distribution. Outliers result from adding extreme values at one end of the distribution. Spikes emerge when adding points with the same value resulting in modes somewhere in the middle of the distribution. As Correll et al. \cite{Correll2019} noted, these types of structures can be identified solely by visual inspection. They are not based on domain-specific information and the definition of severity measures is possible. Since gaps, spikes and outliers are generated using erroneous points, we call these structures data flaws.

\subsection{Quantitative Comparison of Binning Techniques} \label{subsec:quantitativeComparison}
To statistically evaluate the quality of the binning techniques, we use a similarity metric for clustering, the silhouette coefficient \cite{PalacioNino2019}. As with clustering techniques, a binning metric can be considered satisfactory if it has a high degree of separation between bins and a high degree of cohesion within bins \cite{PalacioNino2019}. The silhouette coefficient is one of the most popular approaches that unites the metrics of cohesion and separation in a single measure. The silhouette coefficient is defined in the interval $[-1,1]$. Positive values indicate clear separation between bins, while negative values imply that bins overlap and optimal bin boundaries have not been established by the binning method. A value close to $0$ means that no optimal separation of data points into bins is possible, since the data points are evenly distributed in space. Our hypothesis has been that adaptive binning will result in a partitioning of data points that minimizes within-bin variance and maximizes between-bin variance. Thus, adaptive binning techniques will produce a silhouette coefficient closer to one than the silhouette coefficients calculated for uniform binning. We performed an analysis of variance (ANOVA) on the silhouette coefficients of $72$ distributions, $24$ distributions of each size category. We found a statistically significant difference in the binning quality of the three methods ($p < 0.05$). NB performed best, with an average silhouette coefficient of $0.84$ and a low variance of $0.01$. BB ranked second with an average silhouette coefficient of $0.78$ and a variance of $0.02$. The worst results were achieved by uniform binning, showing an average silhouette coefficient of $0.76$ and a variance of $0.07$. This confirms our hypothesis that adaptive binning methods provide better results.

\subsection{User Study}
The goal of our user study was to address the following research questions: Which binning method represents the underlying data distribution in a visually accurate, yet easily perceivable way? With which visual encoding can univariate distributions be compared more effectively? 

\subsubsection{Hypotheses generation} 
To investigate our research questions, we partitioned our evaluation into two independent sections. The first part, called \textit{Identification Challenge}, focuses on exploring how well, i.e., how correctly, identification tasks of distributions can be performed by observers after binning methods have been applied to the distributions. In this part the distributions are visualized exclusively via the \textit{Bin layout}. The second section of our evaluation, referred to as \textit{Comparison Challenge}, aims to determine which layout of AccuStripes facilitates the visual comparison of multiple distributions. Therefore we evaluate our three layout designs and a line chart design. We formulated the following hypotheses, one per challenge:

\textbf{[H1]} AccuStripes based on adaptive binning methods, i.e., NB or BB binning, will visualize a distribution so that participants will recognize the overall shape of the distribution less error-prone and with a higher confidence compared to AccuStripes using uniform binning.

\textbf{[H2]} Comparing distributions will be faster and less error-prone for participants examining one of our design layouts than examining a line chart representation. According to our definition, a line chart representation of a distribution uses only one line to encode the shape of a distribution \cite{holtz_2011}.

We base hypothesis \textbf{H1} on the observation that NB and BB binning both performed better than uniform binning in the statistical evaluation. We assume that a binning method that provides a high cohesion within bins, i.e., high silhouette coefficient, will bin the data more accurately than another method having a lower cohesion within bins. A more precise binning should make it easier for observers to correctly interpret the underlying distribution, since peaks, gaps, etc. should be less blended together in the resulting binning. Thus, we expect that the more accurate binning of NB and BB binning will also be perceived by the observers. The second hypothesis \textbf{H2} is based on the evaluation of Gogolou et al. \cite{Gogolou2019}, which showed that color coding promotes the perception of higher-level patterns, such as peaks.

\subsubsection{Study Design}
Our study is a judgemental study and similarly in design to the one by Sahann et al.~\cite{Sahann2021}. The study was conducted through a web-based questionnaire implemented in Apps Script \cite{google_2009}. We mailed links of the questionnaires to the participants. They had three days to contribute to the study with their own devices, i.e., standard monitors used at the office. The evaluation was anonymous and there was no way to trace responses back to participants. In the questionnaire, we first gave participants an overview of the study and the objective. We explained AccuStripes and provided a brief insight into the different binning techniques. Second, we asked participants to answer some profile questions, e.g., their gender, age, whether they have experience with analyzing distributions, and whether they change default parameters, e.g., the number of bins, of charts. In total, participants had to answer $72$ multiple-choice questions, $36$ for each section, all with only one correct possibility. The predefined questions were randomized for each participant and the duration of answering each question was recorded. In the \textit{Identification Challenge}, we prepared four questions for each binning type and for each data size category, resulting in $12$ questions per binning method. For each multiple-choice question we showed a distribution aggregated by a particular binning method and visualized it via the \textit{Bin layout}. Three responses were presented to the participants, each of which displayed a distribution in a line chart representation. Participants were asked to select the line graph representation that was visualized in the AccuStripes. In the \textit{Comparison Challenge}, we prepared three questions for each layout design and data size category, resulting in $9$ questions per layout. Each question consisted of six distributions all plotted in the same layout. The first distribution was the ground truth, and the other five were the answer possibilities. To generate the most similar distribution to the ground truth, we first generated the ground truth. Then we added erroneous data points in the extent of $5\%$ to $20\%$ to the original data points of the ground truth. All other distributions were generated randomly. Participants were asked to identify the distribution that they considered most closely resembling the ground truth distribution. In both challenges, each multiple-choice question was accompanied by a supplementary question based on a four-point Likert scale, ranging from "very unconfident" to "very confident". By answering this question, participants could indicate how certain they were with their answer.

\subsection{Results}
A total of $15$ participants joined our study. The participants originated from our working environment of materials researchers. The experiences of our participants with charts or graphs were manifold, but the majority had at least some experience or is working with them on a regular basis. Also the majority of the participants changes default parameters of charts, as the number of bins, sometimes.

For the evaluation of the \textit{Identification Challenge}, we analyzed all questions for correctness, which resulted in a statistically significant result ($p < 0.05$). We found that the binning with the largest number of incorrectly answered questions was BB binning. Of $180$ assessed questions on BB, only $159$ ($88\%$) were answered correctly. NB and uniform binning performed equally good. Of $180$ questions, $170$ ($94\%$) questions were answered correctly for each binning. To find out how confident participants were in answering questions about a particular binning method, we analysed the supplementary questions about their confidence. NB binning performed as well as uniform binning, but the confidence values of the participants for NB binning were slightly higher $(2.28)$ than the confidence values of uniform binning $(2.24)$. The time for answering the questions of uniform binning was fastest with an average of $17.7$ seconds. Questions of NB binning could be answered in an average of $19.7$ seconds. To summarize, we can partially support our hypothesis \textbf{H1}. Uniform binning and NB binning achieved the same correctness scores. However, the confidence value for NB binning is slightly higher, but we could not reach statistical significance for confidence. Although NB binning determined the most accurate binning in the quantitative comparison in \autoref{subsec:quantitativeComparison}, participants could not identify distributions accumulated with NB binning with greater accuracy than distributions accumulated with uniform binning.

In the \textit{Comparison Challenge} the \textit{Bin + Curve layout} ($79\%$) as well as the \textit{Filled Curve layout} ($79\%$) both outperformed the \textit{Bin layout} ($68\%$) in their correctness scores. The line chart representation performed worst for comparing distributions ($55\%$). The participants had the highest confidence for the \textit{Bin + Curve layout} ($2.02$) followed by the \textit{Curve layout} ($1.92$), \textit{Bin layout} ($1.90$) and finally the line chart representation ($1.89$). Answering questions of the \textit{Bin + Curve layout} took participants on average $19.2$ seconds, for the \textit{Filled Curve layout} $19.6$ seconds, for the \textit{Bin layout} $21.9$ seconds. Participants were fastest to answer questions regarding the line chart representation with $17.25$ seconds.
To sum up, our hypothesis \textbf{H2} is supported: There is significant evidence ($p < 0.01$) that each of our layout strategies outperforms the line chart representation. 


\section{Discussion, Limitations and Future Work} \label{sec:Discussion}
Grounded on our user study, we can assume that adaptive binning is as comprehensible to the observer as uniform binning. Therefore, we are interested in comparing the qualities of binning techniques in reducing visual mirages \cite{McNutt2020}. Using adaptive binning, we aim to minimize visualization mirages that lead to inadequate representations of data flaws, as outliers, gaps, or spikes. In our work, we did not measure the detectability of data flaws in AccuStripes through a user study, but we plan this as future work. In the following section, we give a first impression of how uniform, BB and NB binning handle data flaws with AccuStripes.

\subsection{The Representation of Data Flaws in AccuStripes}
\begin{figure*}[h!tb]
  \centering
  \mbox{} \hfill
   \includegraphics[width=\linewidth]{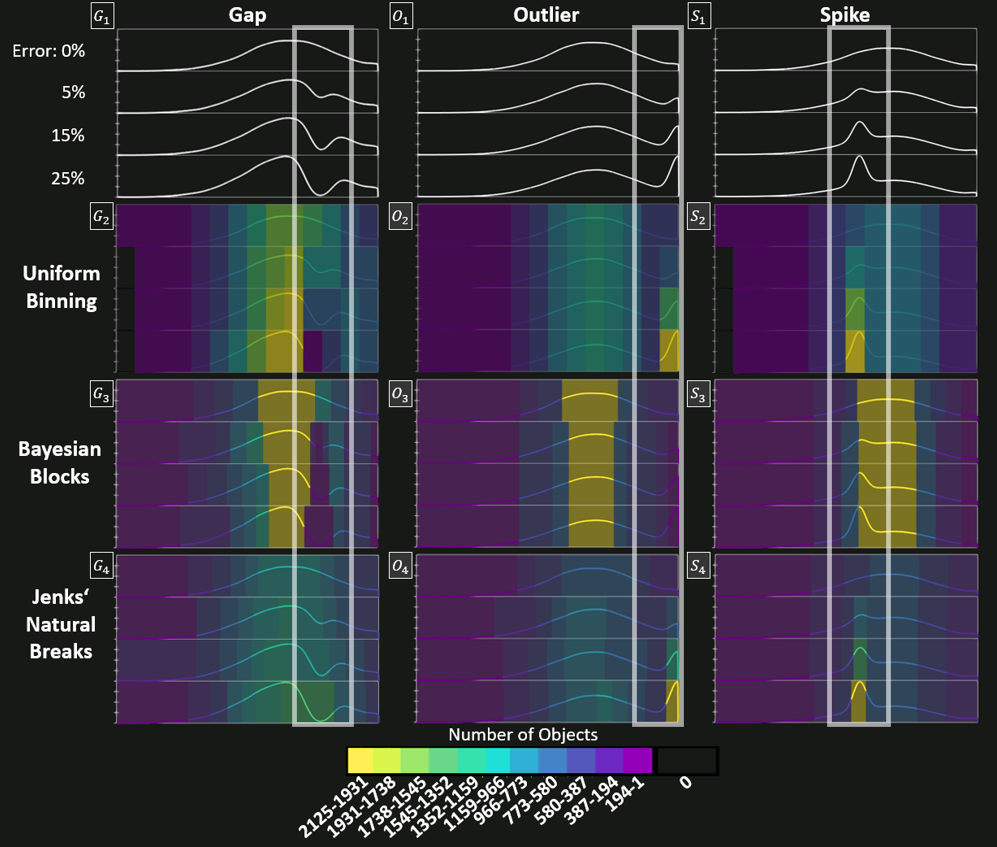}
  \hfill \mbox{}
  \caption{\label{fig:error}%
     Preliminary investigation how uniform, BB, and NB binning visually deal with data flaws, i.e., gaps, outliers and spikes, in a distribution: $G_{1}$ shows four distributions with $0\%$, $5\%$, $15\%$, and $25\%$ erroneous points leading to a varying gap sizes. $G_{2}$, $G_{3}$ and $G_{4}$ demonstrate how the different binning strategies represent the gap through their binning and coloring. All binning methods visualize gaps clearly visible. $O_{1}$ demonstrates four distributions with outliers of varying severity. Uniform ($O_{2}$) and NB binning ($O_{4}$), both represent the outliers through color coding. BB binning ($O_{3}$) does not show changes in binning or coloring after more than $5\%$ of flawed points are present in the distribution. $S_{1}$ displays the distributions with spikes in varying size. Again, uniform ($S_{2}$) and NB binning ($S_{4}$) indicate the spikes through their binning and coloring. BB binning ($S_{3}$) shows no change above $5\%$ of faulty points.}
\end{figure*}

To give an idea of how uniform and adaptive binning represent data flaws in AccuStripes, we generated distributions of the size $\num{10000}$. First, we produced a Gaussian distribution with $0\%$ flawed points. Next, we created three distributions per flaw type, each based on the original Gaussian distribution, but modified by $5\%$, $15\%$, and $25\%$ erroneous points of the respective flaw type. The distributions indicating a gap with varying severity are shown in \autoref{fig:error}$G_{1}$. \autoref{fig:error}$O_{1}$ demonstrates the distributions illustrating outliers. \autoref{fig:error}$S_{1}$ displays the datasets representing spikes with an increasing number of flawed points. \autoref{fig:error} overviews our erroneous distributions and how the different binning methods represent the three flaw types. In each representation the ordering of the distributions is $0\%$, $5\%$, $15\%$, and $25\%$. We have highlighted the erroneous positions in the distributions with a white frame to facilitate their identification. Comparing the location of the gap over the various binning strategies in \autoref{fig:error}$G_{2}$, \autoref{fig:error}$G_{3}$, and \autoref{fig:error}$G_{4}$, we can get a first impression how the methods deal with missing points. A difference in binning and coloring can be perceived in all three binning methods for each increase in flawed points. \autoref{fig:error}$O_{2}$, \autoref{fig:error}$O_{3}$, and \autoref{fig:error}$O_{4}$, indicate how the binning methods behave with an increase in outliers. We can observe that both uniform and NB binning reflect the higher number of extreme values in their coloring for each increase of erroneous points. BB binning reveals a change in binning and color coding only from $0\%$ to $5\%$. Thereafter, an increase in outliers can no longer be recognized in \autoref{fig:error}$O_{3}$. Finally, \autoref{fig:error}$S_{2}$, \autoref{fig:error}$S_{3}$, and \autoref{fig:error}$S_{4}$ illustrate how the binning techniques handle a spike of varying severity. Once more, uniform and NB binning clearly showcase each increase of erroneous points. BB binning only shows a change in grouping and color coding from $0\%$ to $5\%$. No change is visible in the subsequent stages in \autoref{fig:error}$S_{3}$. BB binning aims to group segments of the distribution in which the data points' values remain the same. For this reason, segments that have a high variation are divided into many small bins. In the cases of outliers shown in \autoref{fig:error}$O_{3}$ and of spikes in \autoref{fig:error}$S_{3}$, a strong increase in data points has to be accumulated at a locally very restricted position. Thus, many narrow bins are generated which contain only a small number of data points. In the presented \textit{Bin + Curve layout}, bin boundaries are not explicitly drawn, so many narrow adjacent bins with the same color coding are not clearly visible. 

\subsection{Color Schemes and Number of Bins}
In AccuStripes, the choice of the color scheme significantly affects the interpretability of the distributions. For the user study we opted for \textit{Viridis}, but we conducted another user study in advance using the color scheme \textit{Bodyheat} \cite{Reda2018}. In this case we visualized small distributions consisting of $100$ data points. \textit{Bodyheat} is a sequential colormap with a monotonically increasing luminance of a limited number of hues. Although this color scheme is perceptually effective, Mittelstaedt et al. \cite{Mittelstaedt2014} showed that multi-hue colormaps outperform other mappings, since they have a very low probability of producing the contrast effect, which makes pixels appear brighter if surrounded by a darker area. AccuStripes are a representation visually similar to heatmaps, which are susceptible to suffer from the contrast effect \cite{Han2020a}. Thus, we chose to use multi-hue colormaps. Beside \textit{Viridis}, other multi-hue colormaps are \textit{Magma} and \textit{Plasma}. These color schemes are also formed by tracing curves through a perceptually uniform color model and have the same properties as \textit{Viridis}. While evaluating error rate and time for judgement tasks, Liu et al. \cite{Liu2018} discovered that \textit{Magma} and \textit{Plasma} performed inferior to \textit{Viridis}. A sequential colormap resembling \textit{Viridis} is \textit{YlGnBu} by ColorBrewer \cite{Brewer2003}, with hue and luminance trajectories being very similar and only differing chroma trajectories \cite{Zeileis2020}. For \textit{YlGnBu} light colors have a low chroma, while they have an increasing chroma in \textit{Viridis}. As a result, the colors of \textit{Viridis} have a high chroma value throughout, which does not work well on a bright background, as all the shaded areas convey a high intensity. Thus, \textit{Viridis} works better on dark background. As defined by Wang et al. \cite{Wang2008} the separation of foreground and background works best if the foreground color is bright and highly saturated, while the background color should be de-saturated. This statement supports our decision to select a dark background. In our study we could not focus on the effect that different color schemes have on the perception of the AccuStripes. More research is required to provide accurate information in this regard.

Furthermore, the discrete color schemes used for AccuStripes consisted of 12 colors. Since adaptive binning techniques can produce narrow bins, a reduction to ten colors could be beneficial. Fewer colors allow for greater differences in the colors, so identifying especially small regions should be easier for observers. The overall number of bins depends on the binning technique used and on the shape of the distribution. Multi-modal distributions with many sharp spikes will lead to a high number of narrow bins for BB and NB binning. Distributions with few modes that just vary slightly in their height result in very few and wide bins for both adaptive binning methods. NB and BB binning did not produce significantly higher numbers of bins than uniform binning based on the Sturge's Rule. Sahann et al. \cite{Sahann2021} argue, that $\sim20$ bins are sufficient to accurately perceive distributions. We focused on the evaluation of AccuStripes using types of Gaussian distributions. Further tests are necessary to provide insights how AccuStripes represent highly alternating multi-modal distributions. Furthermore, we only examined two of a variety of existing adaptive binning techniques. Since AccuStripes are intended for non-statisticians, accurate binning of the data should be obtainable without knowledge of the technique and adjustment of parameters. One possible adaptive method we plan to explore next is the clustering method DBScan \cite{Schubert2017a}.

\section{Conclusion} \label{sec:Conclusion}
In this paper we introduced the visual metaphor AccuStripes to represent a distribution of data. Our goal has been to accumulate a distribution in an accurate but easy-to-interpret form and present its shape using color. To ease the comparison of many distributions AccuStripes were aligned in a stacked fashion and we explored three different layout strategies, the \textit{Bin layout}, the \textit{Bin + Curve layout} and the \textit{Filled Curve layout}. Our user study showed that the adaptive binning technique NB performed as well as uniform binning. This suggests that adaptive binning could be an alternative for accumulating distributions compared to uniform binning based on the Sturge's rule. More research is necessary to determine whether adaptive binning provides more accurate binning of distributions than uniform binning based on other rules of thumb. Our user study on the most efficient layout in AccuStripes for comparing distributions yielded a clear result: Colored representations of the distributions facilitate the comparison. The line chart representation performed worst, having the largest number of incorrectly answered questions. Of our proposed layouts, the \textit{Bin layout} received the best feedback. This shows that the combination of stripes and lines as visual encoding allows for the most correct comparison of distributions. The line revealed the shape of the distribution, while the stripes in the background prominently showed the color pattern.

\FloatBarrier

\bibliographystyle{abbrv-doi}

\bibliography{Papers}
\end{document}